\documentclass[twocolumn]{aastex63}
\usepackage{CJK}
\usepackage{amsmath}
\usepackage{verbatim}
\definecolor{blue}{RGB}{50, 80, 255}

\newcommand\aastex{AAS\TeX}

\usepackage{hyperref}
\definecolor{blue}{RGB}{50, 80, 255}
\definecolor{red}{RGB}{255, 50, 50}

\submitjournal{ApJ}

\shorttitle{\aastex\ IR Excesses around Bright White Dwarfs from Gaia and unWISE I}
\shortauthors{Xu, Lai, \& Dennihy}

\begin{document}
\begin{CJK}{UTF8}{gbsn}
\title{Infrared Excesses around Bright White Dwarfs from Gaia and unWISE I}

\correspondingauthor{Siyi Xu} 
\email{sxu@gemini.edu}

\author[0000-0002-8808-4282]{Siyi Xu (许\CJKfamily{bsmi}偲\CJKfamily{gbsn}艺)}
\affil{NSF's NOIRLab/Gemini Observatory, 670 N. A'ohoku Place, Hilo, Hawaii, 96720, USA}

\author[0000-0001-9372-4611]{Samuel Lai (赖民希)}
\affil{NSF's NOIRLab/Gemini Observatory, 670 N. A'ohoku Place, Hilo, Hawaii, 96720, USA}

\author[0000-0003-2852-268X]{Erik Dennihy}
\affil{NSF's NOIRLab/Gemini Observatory, Casilla 603, La Serena, Chile}

\begin{abstract}

Studies of excess infrared radiation around white dwarfs provide important constraints on the evolution of planetary systems and low-mass stars beyond the main sequence stage. In this paper series, we focus on identifying and characterizing bright white dwarfs with an infrared excess. Here, we present 188 infrared excess candidates from {\it Gaia} and unWISE, 147 of which are new discoveries. Further characterization of this sample can significantly increase the current list of white dwarf debris disks and white dwarfs with low-mass companions.

\end{abstract}


\section{Introduction} \label{sec:intro}

White dwarfs are the most common end point of stellar evolution. Because of their relatively high temperatures, the fluxes peak in the ultraviolet/optical, making them ideal objects to search for infrared excesses. Infrared excesses around white dwarfs can come from any objects that are cooler than the white dwarf, such as a cooler white dwarf, an M dwarf, a brown dwarf, or a debris disk. The last two cases are the focus of this work. The first L dwarf was discovered around the white dwarf GD~165 \citep{BecklinZuckerman1988}. Around the same time, another white dwarf G~29-38 was discovered to display excess infrared radiation \citep{ZuckermanBecklin1987}, later recognized to be from a circumstellar dust disk \citep{Tokunaga1990, Graham1990b}. The next discovery of a dust disk came 18 years later around GD~362 \citep{Becklin2005,Kilic2005}. With the launch of the {\it Spitzer Space Telescope}, more than 50 white dwarfs with an infrared excess from a debris disk have been identified \citep{Farihi2016, Dennihy2020}.

The standard model is that these disks are remnants of extrasolar minor planets that were perturbed into the tidal radius of the white dwarf \citep{DebesSigurdsson2002, Jura2003}. Afterwards, these disks might evolve passively under the effects of the Poynting-Robertson drag \citep{Rafikov2011a} or they might experience collisional cascades as more disrupted material continuously arrives at the disk \citep{KenyonBromley2017a}. The nominal frequency of white dwarf debris disks is 2-4\% \citep{Barber2014, Rocchetto2015, Wilson2019}. These dust disks deposit material onto the white dwarf's surface, where one can determine the chemical compositions of the accreting extrasolar planetary material \citep{JuraYoung2014}. While atmospheric pollution is detected in 30\% of white dwarfs \citep{Zuckerman2003, Zuckerman2010, Koester2014a}, dusty white dwarfs are prime targets for such studies because their atmospheres are often heavily polluted \cite[e.g.,][]{Xu2019b}. Some dusty white dwarfs also display significant amounts of circumstellar gas \cite[e.g.,][]{Gaensicke2006}. In addition, observational evidence from actively disintegrating asteroids has been discovered around at least two dusty white dwarfs \citep{Vanderburg2015, Manser2019}. Every single dusty white dwarf has a story of its own and provides a laboratory to characterize properties of extreme planetary systems.

Detached white dwarf-brown dwarf pairs can also display infrared excesses. There are only nine such systems known to date and their occurrence rate is estimated to be 0.5-2.0\% \citep{Girven2011,Steele2011}. Studies of white dwarf-brown dwarf pairs allow us to investigate binary formation and evolution with extreme mass ratios \citep{Rappaport2017b, Longstaff2019}. In addition, the atmosphere of the brown dwarf is often strongly irradiated by the white dwarf, making it a good hot Jupiter analog \citep{TanShowman2020}.

The {\it Spitzer Space Telescope} played a major role in identifying infrared excesses around white dwarfs \citep{Chen2020}. However, most {\it Spitzer} searches have been focused on white dwarfs that are known to be polluted \citep[e.g.,][]{Farihi2009, XuJura2012} or in a specific temperature range \citep{Rocchetto2015, Wilson2019}. There have been significant efforts to search for infrared excesses around white dwarfs with data from the {\it Wide-field Infrared Survey Explorer} ({\it WISE}, \citealt{Debes2011b,Hoard2013,Dennihy2016,Dennihy2017,Rebassa-Mansergas2019}). The main source of false positives is contamination from source confusion due to the 6{\farcs}0 {\it WISE} beam size. For example, \citet{Barber2014} followed up on 16 {\it WISE}-selected dusty white dwarf candidates and found only four systems are free from contamination in J and H band. \citet{Dennihy2020} followed up on a sample of 22 {\it WISE}-selected candidates that are also clean in deep ground-based JHK imaging. They found that the {\it Spitzer} images of eight systems are still confused with nearby objects and concluded that confusion is unavoidable for {\it WISE}-selected infrared excesses. This will be the main limitation for this study as well.

The recent {\it Gaia Data Release 2} has returned $\simeq$ 260,000 high confidence white dwarfs \citep{Gentile-Fusillo2019}. Combined with the newly released unWISE catalog \citep{Schlafly2019}, this provides a unique opportunity to search for and characterize infrared excesses around a large sample of white dwarfs. That is the overarching goal of this paper series. In this work, we present our initial selection of infrared excess candidates from {\it Gaia} and unWISE. We start with visiting the current sample of white dwarfs with infrared excesses in Section~\ref{sec:CurrentSample}. Our selection criteria and new infrared excess candidates are presented in Section~\ref{sec:NewSample}. We assess our selection criteria and the properties of white dwarfs with infrared excesses in Section~\ref{sec:discuss} and our conclusions are given in Section~\ref{sec:conclusion}.

\section{Current Sample of White Dwarfs with Infrared Excesses \label{sec:CurrentSample}}

In total, there are over 50 white dwarfs with a {\it Spitzer}-confirmed, dusty infrared excess \citep{Farihi2016, Dennihy2020}. Here, we focus on the 40 bright systems listed in Table~\ref{tab:knownir}, which have {\it Gaia} G $<$ 17.0 mag and are therefore better suited for further characterization. Pollution is often detected in the atmospheres of white dwarfs with debris disks \cite[e.g.,][]{Xu2019b},  i.e., Z in the spectral type (SpT) column, corroborating the picture that the white dwarf is accreting from the circumstellar material. For the white dwarf-brown dwarf pairs, there are often additional observational signatures, such as radial velocity variations and photometric variations, that confirmed the binary nature of these systems \citep[e.g.,][]{Maxted2006,Casewell2015}.

\begin{deluxetable*}{lcccccccccccc}
\tablecaption{ \label{tab:knownir} Known bright white dwarfs (G $<$ 17.0 mag) with an infrared excess. The RA, DEC, and G magnitude are taken from the {\it Gaia Data Release 2} \citep{GaiaDR2}. To be consistent with the rest of the {\it Gaia} sample, the white dwarf parameters (T$_\mathrm{eff}$ and log g) are all H-atmosphere fits from \cite{Gentile-Fusillo2019}, which are derived using the {\it Gaia} photometry and parallax. Exs is the assessment of the infrared excess based on the methods presented in this paper (Section~\ref{sec:method}). Mag and color means the infrared excess is identified from the \textit{W1}\,\&\,\textit{W2} magnitude excess or the (\textit{W1}-\textit{W2}) color excess, respectively. N means there is no unWISE photometry while ... means no infrared excess is identified. For dusty white dwarfs, we list the reference where the {\it Spitzer} observations are reported. Most white dwarf-brown dwarf pairs do not have {\it Spitzer} observations, so we list the discovery paper. }
\tablehead{
\colhead{Name} & \colhead{RA} & \colhead{DEC} & \colhead{G}& \colhead{SpT} &   \colhead{T$_\mathrm{eff}$} & \colhead{log g } & \colhead{Exs} & \colhead{Reference} \\
& (deg) & (deg) & (mag) & &  (K) & (cm s$^{-2}$)
}
\startdata
Dust Disks \\
       G 166-58 &     224.527739 &       29.622289 &  15.5 &      DAZ  &   7333 &  7.98 &  N        &     \citet{Farihi2008b} \\
   WD 2115$-$560 &     319.905524 &      -55.838173 &  14.3 &      DAZ  &   9674 &  7.99 & N          &            \citet{Farihi2009} \\
WD 2221$-$165 &     336.072595 &      -16.263551 &  16.0 &      DAZ &   9867 &  8.10 &    Mag &                        \citet{Farihi2010b}  \\
          GD 362 &     262.893117 &       37.088152 &  16.0 &     DAZB\tablenotemark{a}  &  10513 &  8.17 &       Color, Mag   &                      \citet{Farihi2008a}  \\
   WD 1541+650 &     235.437169 &       64.897776 &  15.6 &      DAZ  &  11278 &  8.00 &  Color, Mag &                          \citet{Kilic2012} \\
   WD 0307+077 &      47.537841 &        7.958517 &  16.2 &      DAZ  &  11351 &  8.04 &      ...    &                         \citet{Farihi2010b} \\
        G 29-38 &     352.196764 &        5.247250 &  13.1 &      DAZ  &  11357 &  8.02 &      N    &                         \citet{Reach2005a}  \\
         GD 16 &      27.237675 &       19.040493 &  15.6 &     DAZB\tablenotemark{a}  &  11666 &  8.27 &       Color, Mag    &                         \citet{Farihi2009} \\
 EC 21548$-$5908 &     329.599551 &      -58.898160 &  15.8 &      DAZ  &  11688 &  8.01 &  Color, Mag &                       \citet{Dennihy2020} \\
   WD 1150$-$153 &     178.313419 &      -15.610387 &  16.0 &      DAZ  &  11917 &  8.02 &  Color, Mag &  \citet{Jura2009a} \\
   WD 0950$-$572 &     147.914279 &      -57.444636 &  15.0&       DA  &  12127 &  7.86 &   N      &                         \citet{Barber2016} \\
        GD 133 &     169.801230 &        2.342646 &  14.7 &      DAZ  &  12259 &  8.04 &  Color, Mag &                          \citet{Jura2007b} \\
   WD 2132+096 &     323.711472 &        9.922079 &  16.0 &      DAZ  &  13056 &  7.96 &    ...      &                      \citet{Bergfors2014} \\
   WD 1132+470 &   173.702480 &       46.809370 &    16.4 &    DA & 14052 &      8.53 &        ... &                            \citet{Wilson2019}               \\
         GD 40 &      45.720883 &       -1.142875 &  15.5 &     DBAZ\tablenotemark{a}  &  14504 &  8.02 &  Color, Mag &                           \citet{Jura2007b} \\
        GD 685 &      17.387871 &      -19.021850 &  16.2 &      DAZ  &  14516 &  7.96 &     ...     &                       \citet{Dennihy2020} \\
 EC 23379$-$3725 &     355.152476 &      -37.145862 &  16.2 &      DAZ  &  14762 &  7.97 &  Color, Mag &                       \citet{Dennihy2020} \\
         GD 56 &      62.759052 &       -3.973494 &  15.6 &      DAZ  &  15151 &  8.02 &  Color, Mag &                           \citet{Jura2007b} \\
         GD 61 &      69.664087 &       41.158515 &  14.8 &     DBAZ\tablenotemark{a}  &  16034 &  7.98 &    Mag &                        \citet{Farihi2011a} \\
   WD 0106$-$328 &      17.150280 &      -32.628809 &  15.4 &      DAZ  &  16226 &  8.03 &   Color &                         \citet{Farihi2010b} \\
   WD 1349$-$230 &     208.183826 &      -23.334905 &  16.6 &     DBAZ\tablenotemark{a}  &  16597 &  7.86 &    N      &                         \citet{Girven2012} \\
    WD 0842+572 &    131.510132 &     57.057851 &      16.8 &         DA &    16617 &     7.98 &             N &                          \citet{Swan2020} \\
 WD 2329+407 &   352.900095 &    41.024766 &     13.9 &        DA &   16689 &    7.99 &            N &                         \cite{Swan2020} \\
   WD 0110$-$565 &      18.088172 &      -56.241105 &  15.8 &     DBAZ\tablenotemark{a}  &  18249 &  7.91 &  Color, Mag &                         \citet{Girven2012} \\
   WD 2328+107 &     352.673303 &       11.034950 &  15.6 &       DA  &  18482 &  7.67 &   ...       &                      \citet{Rocchetto2015} \\
  WD  0420-731 &   64.907685 &  -73.062304&    15.6 &       DA &  18553 &   7.95 &  Color, Mag &                        \citet{Swan2020} \\
   WD 1015+161 &     154.515450 &       15.865973 &  15.7 &      DAZ  &  18854 &  8.01 &  Color, Mag &                           \citet{Jura2007b} \\
       Ton 345 &     131.413188 &       22.957582 &  15.9 &      DBZ\tablenotemark{a}  &  18889 &  7.87 &  Color, Mag &                    \citet{Brinkworth2012} \\
 EC 01129$-$5223 &      18.754542 &      -52.129252 &  16.5 &       DA  &  19831 &  7.92 &   ...       &                       \citet{Dennihy2020} \\
    WD 1226+110 &     187.249505 &       10.675740 &  16.4 &      DAZ  &  20168 &  8.06 &  Color, Mag &                    \citet{Brinkworth2012} \\
   WD 1018+410 &     155.481238 &       40.837419 &  16.4 &      DAZ  &  20858 &  8.01 &  Color, Mag &                      \citet{Rocchetto2015} \\
 WD 0843+516 &     131.759455 &       51.481078 &  16.1 &      DAZ  &  21053 &  7.79 &  Color, Mag &                        \citet{JuraXu2012} \\
   WD 1929+011 &     292.986917 &        1.295522 &  14.3 &      DAZ  &  21680 &  7.98 &      N    &                      \citet{Rocchetto2015} \\
      WD 0420+520 &      66.065321 &       52.169586 &  15.0 &       DA  &  22563 &  8.07 &  Color, Mag &                         \citet{Barber2016} \\
 EC 05365-4749 &   84.472561 &  -47.968136 &    15.6 &       DA &  22967 &   8.13 &  Color, Mag &                       \citet{Swan2020} \\
   WD 0010+280 &       3.338109 &       28.338806 &  15.7 &      DAZ  &  24607 &  7.77 &    Mag &                             \citet{Xu2015} \\
\hline
Brown Dwarfs \\
   NLTT 5306 &      23.886811 &       14.764879 &  16.9 &  DA+L4-7 &    7729 &  7.57 &    Mag &                        \citet{Steele2013} \\
     GD 1400 &      26.841248 &      -21.947712 &  15.2 &  DA+L6 &     10842 &  7.97 &       Mag   &             \citet{FarihiChristopher2004} \\
 WD 2218$-$271\tablenotemark{b} &    335.350105 &      -26.848189 &    14.8 & DA+???&    11437 &      7.67 &   ... &                                \citet{Wilson2019} &              \\
    WD 0137$-$349 & 24.928335 & -34.711204 & 15.4 & DA+L6-8 & 16396 & 7.47 & Mag & \citet{Maxted2006} \\
  \enddata
\tablenotetext{a}{This white dwarf has a helium-dominated atmosphere and there are more accurate parameters available in the literature.}
\tablenotetext{b}{The origin of the infrared excess is unclear for this system. \citet{Wilson2019} suggested that it is not a dust disk because the white dwarf's atmosphere is not polluted. In this work, we put it into the white dwarf-brown dwarf category.}
\end{deluxetable*}

The current sample of white dwarfs with an infrared excess extends a wide range of parameter space. The debris disks are mostly found around white dwarfs of 25,000 --
9,000~K (cooling age of 30 Myr -- 1 Gyr). At the hot end, the lack of dust disks can be explained as a result of dust sublimation \citep{vonHippel2007}. At the cool end, the lack of dust disks can be attributed to the decreasing numbers of tidal disruption events which lead to the formation of these disks \citep[e.g.,][]{Debes2012a,Veras2016}. However, these are not strict limits and there are systems outside of this temperature range. For example, the recent discovery of a dust disk candidate\footnote{By the definition in this paper, this system is considered a disk candidate because no {\it Spitzer} photometry is available.} around a 3~Gyr old (5790~K) white dwarf shows that older debris disks also exist \citep{Debes2019}. There is no theoretical limit on the temperatures of white dwarfs with brown dwarf companions.

\begin{figure}
\epsscale{1.2}
\plotone{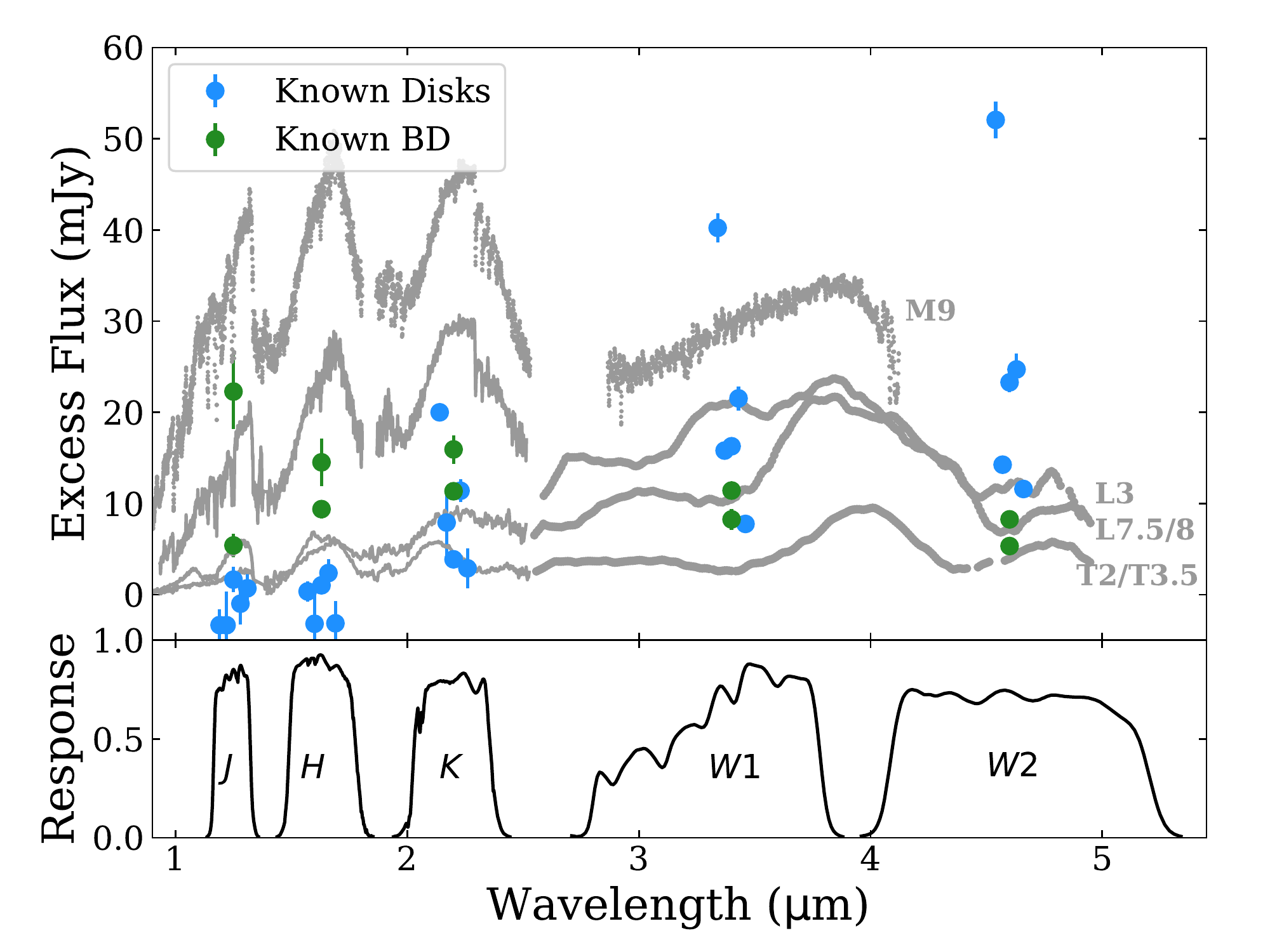}
\caption{Excess infrared fluxes for two white dwarf-brown dwarf pairs (NLTT 5306 and WD 0137$-$349) and five white dwarf debris disks (WD~1541+650, WD~1150$-$153, GD~56, WD~1226+110, and WD~0420+520). For clarity, there are some small offsets in wavelength for the photometry points. Library spectra for low-mass companions are also shown for comparison \citep{Cushing2005, Rayner2009, Stephens2009,SorahanaYamamaura2012}. All data are scaled to a distance of 10~pc. The bottom panel is the response function of each filter. The {\it WISE} fluxes for dust disks and low-mass companions are very similar and often indistinguishable.
}
\label{fig:sed_comp}
\end{figure}

The Spectral Energy Distributions (SED) of white dwarfs with brown dwarf companions and debris disks are very similar between 1-- 5 $\mu$m, as shown in Figure~\ref{fig:sed_comp}.  This was recognized in the early days of infrared excess searches around white dwarfs. In fact, the discovery paper on G 29-38 suggested that the infrared excess could come from either a brown dwarf companion or a dust disk \citep{ZuckermanBecklin1987}. More recently, WD~J1557+0916 was first identified as a white dwarf with a dust disk from the {\it Spitzer} observations \citep{Farihi2012a} but later it became clear that the infrared excess actually comes from a brown dwarf companion and a circumbinary disk \citep{Farihi2016}. We caution that without additional follow-up, the origin of the infrared excess can be ambiguous. Nevertheless, we can use the current sample of known white dwarfs with infrared excesses to guide our search for new systems.

\section{Identifying New Infrared Excesses \label{sec:NewSample}}

\subsection{Initial Sample \label{sec:sample}}

We start with white dwarf candidates identified from the {\it Gaia Data Release 2} reported in \citet{Gentile-Fusillo2019}. We focus on bright white dwarfs with (i) {\it Gaia} G $<$ 17.0~mag, which are more likely to have detections at the {\it WISE} bands and well suited for further characterization. We also require (ii) the white dwarf probability P$_\mathrm{WD}$ $>$ 0.75, which has an estimated false positive rate of 4\%. \citet{Gentile-Fusillo2019} reported white dwarf parameters (e.g., effective temperature T$_\mathrm{eff}$ and surface gravity log~g) using the {\it Gaia} photometry and parallaxes and we use results from the H-atmosphere fits for the rest of the paper. We also limit the sample to (iii) T$_\mathrm{eff}$ $<$ 50,000~K and log~g $>$ 7.0~cm~s$^{-2}$. The temperature limit is to exclude hot white dwarfs and pre-white dwarfs in planetary nebulae. The surface gravity limit is to exclude unresolved white dwarf and main sequence binaries, which can appear to have a small surface gravity for photometrically determined parameters. There are 6002 high probability white dwarf candidates in our starting sample, which also includes 36 known dusty white dwarfs and 4 white dwarfs with low-mass companions in Table~\ref{tab:knownir}.

\subsection{ALLWISE vs unWISE}

{\it WISE} photometry is crucial in identifying white dwarfs with infrared excesses because the excess flux is most apparent in the first two {\it WISE} bands  (see Figure~\ref{fig:sed_comp}). Previously, the WISE Preliminary Release Catalog, the WISE All Sky Survey, and the ALLWISE data release have been used to search for infrared excesses \citep{Debes2011b, Hoard2013, Dennihy2016, Dennihy2017, Rebassa-Mansergas2019}. Recently, there is a new data release called the unWISE catalog, which combines five years (i.e., 2010, 2014--2017) of {\it WISE} images at 3.4~$\mu$m and 4.6~$\mu$m \citep{Lang2014,Meisner2017,Schlafly2019}. The exposure time of the unWISE co-adds is about a factor of five of the ALLWISE co-adds \citep{Cutri2013}, leading to 0.7~mag fainter detection limits at 5$\sigma$. In addition, thanks to the improved modeling using the forced photometry algorithm and the {\it crowdsource} photometry pipeline \citep{Schlafly2018}, the unWISE photometry is more reliable, particularly in crowded regions.

\begin{figure*}
\gridline{\fig{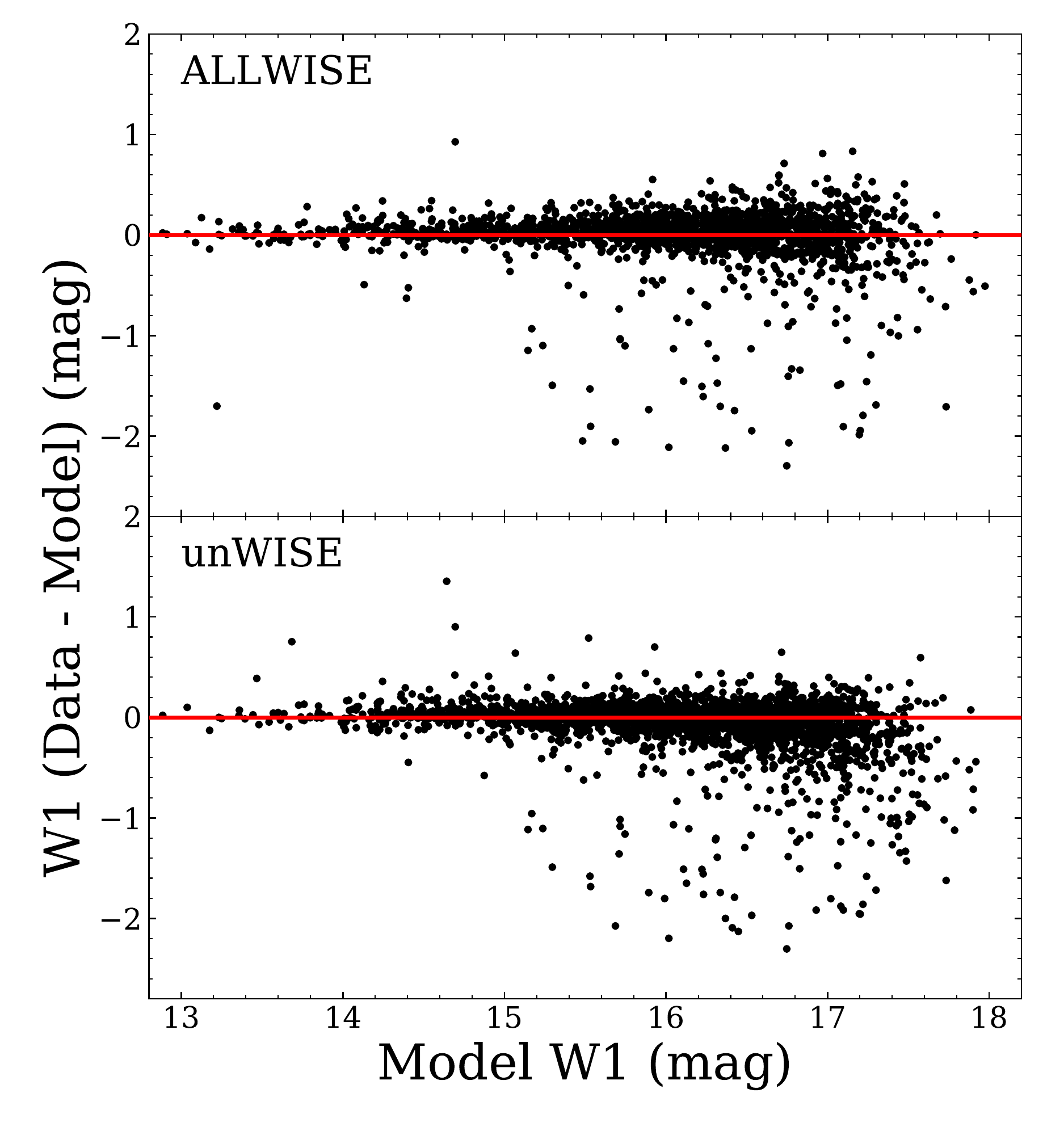}{0.45\textwidth}{}
\fig{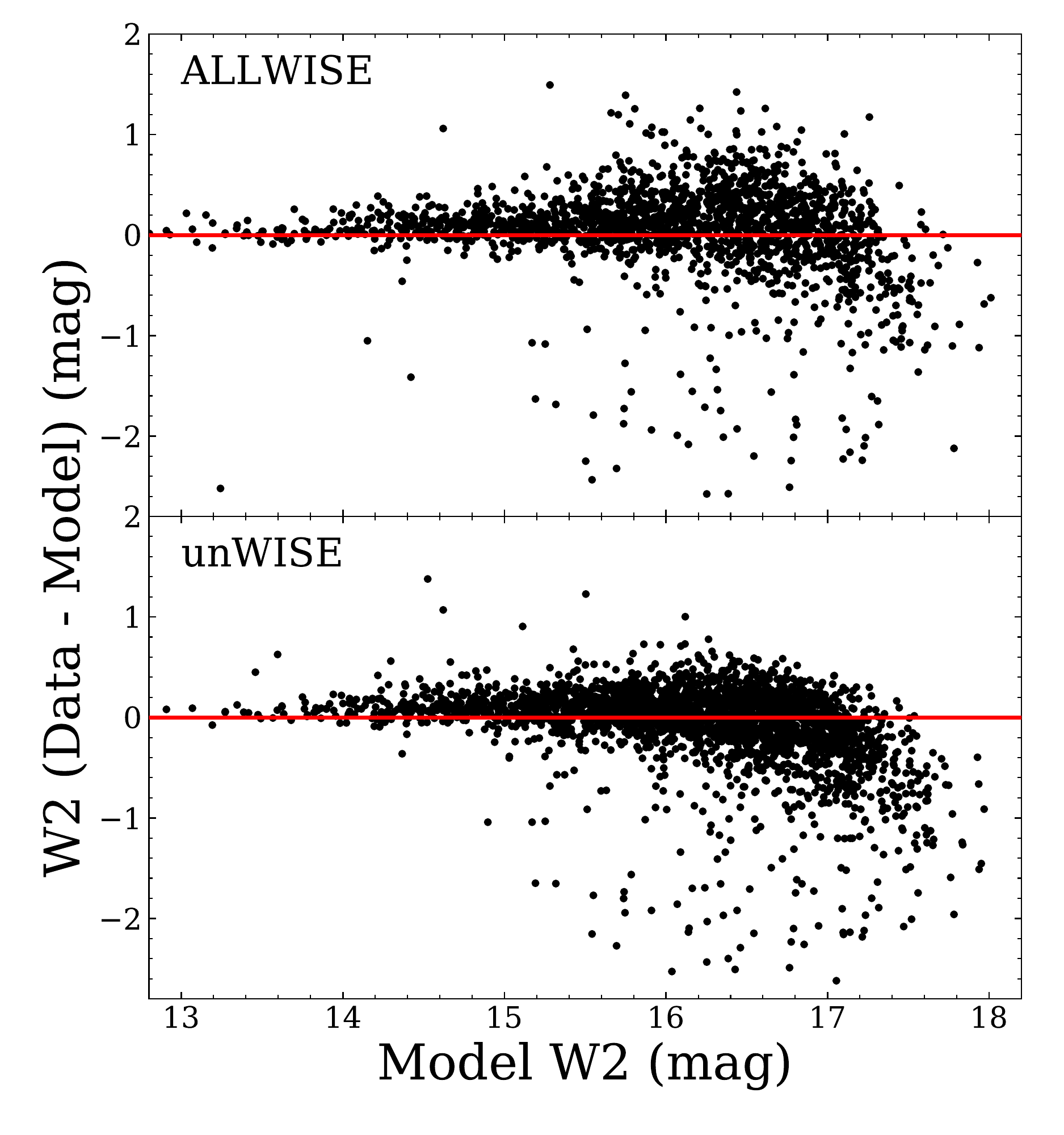}{0.45\textwidth}{}
}
\caption{A comparison between ALLWISE photometry, unWISE photometry, and the expected single white dwarf magnitudes from models. The objects with a negative magnitude difference are mostly infrared excess candidates. In general, there is a good agreement for bright objects between ALLWISE, unWISE and the model magnitudes; however, for faint objects, the scatter becomes much larger between the ALLWISE and the model magnitudes. The downward trend at the faint end (e.g., W2 $>$ 17.0) is due to the sensitivity limit of the survey.
}
\label{fig:comp_mag}
\end{figure*}

To have an independent assessment of the unWISE catalog, we cross-correlated the 6002 {\it Gaia} white dwarfs with unWISE and ALLWISE separately. unWISE returned 2886 reliable detections in both {\it W1} and {\it W2} while ALLWISE only returned 1858 detections\footnote{Reliable photometry means that the measured position in the WISE catalog agrees to within 3$\sigma$ of the calculated position from {\it Gaia} and the white dwarf is the only source detected within a 3{\farcs}0 radius in different catalogs. See description for Steps I and III in Section~\ref{sec:method}.}. For the unWISE photometry, we added an additional 3\% systematic uncertainty in quadrature to the reported statistical uncertainty to account for the fluctuation of the zero points throughout the unWISE coadds \citep{Lang2014, Meisner2017}. A comparison between the unWISE photometry and ALLWISE photometry with the expected single H-atmosphere white dwarf photometry from models (P. Bergeron, private communications) is shown in Figure~\ref{fig:comp_mag}. At the bright end, the ALLWISE and unWISE performances are similar. At the faint end, the unWISE magnitudes have a much smaller scatter compared to those of ALLWISE, thanks to the better photometry algorithm of unWISE \citep{Schlafly2018}. However, the faintest objects are at the sensitivity limit of unWISE and our assessment of their infrared excess might not be reliable.

We also calculated the median uncertainty in each magnitude bin for ALLWISE and unWISE respectively in Table~\ref{tab:unwise}. On the bright end ($<$ 16.0 mag), the uncertainties are comparable between ALLWISE and unWISE. On the faint end ($>$ 16.0 mag), the unWISE uncertainties are significantly smaller, particularly at \textit{W2}. 

\begin{deluxetable}{cccccc} 
\tablecaption{Median Uncertainties from ALLWISE \& unWISE \label{tab:unwise} }
\tablewidth{0pt}
\tablehead{
& \colhead{ALLWISE} & \colhead{unWISE} & \colhead{ALLWISE} & \colhead{unWISE} 
\\
 \colhead{Mag} & \colhead{ \textit{W1} (mag)} & \colhead{ \textit{W1} (mag)} &  \colhead{ \textit{W2}  (mag)} & \colhead{ \textit{W2}  (mag)}
 }
\startdata
$<$ 14.0 	& 0.025 & 0.037 & 0.032 & 0.043\\
14.0 -- 15.0 & 0.030 & 0.040 & 0.055 & 0.055  \\
15.0 -- 16.0 & 0.042 & 0.048 & 0.108 & 0.086\\
16.0 -- 17.0 & 0.069 & 0.062 & 0.232 & 0.143\\
$>$ 17.0 & 0.120 & 0.084 & 0.426 & 0.215\\
\enddata
\end{deluxetable}

unWISE is a static-sky catalog and it did not account for proper motion -- one major drawback compared to ALLWISE \citep{Schlafly2019}. Fortunately, for the 6002 white dwarfs of interest, 5910 systems (98.5\%) move less than four arcseconds over the seven year time period of the unWISE co-adds. Four arcseconds is about half of the {\it WISE} beam size, which is also our crossmatch search radius. Therefore, proper motion is not a concern for most white dwarfs in this sample. Given unWISE's smaller uncertainty, larger number of detections, and more reliable photometry for faint objects, we decided to use it for our analysis.

\subsection{Selection Criteria \label{sec:method}}

\begin{figure}
\epsscale{1.2}
\plotone{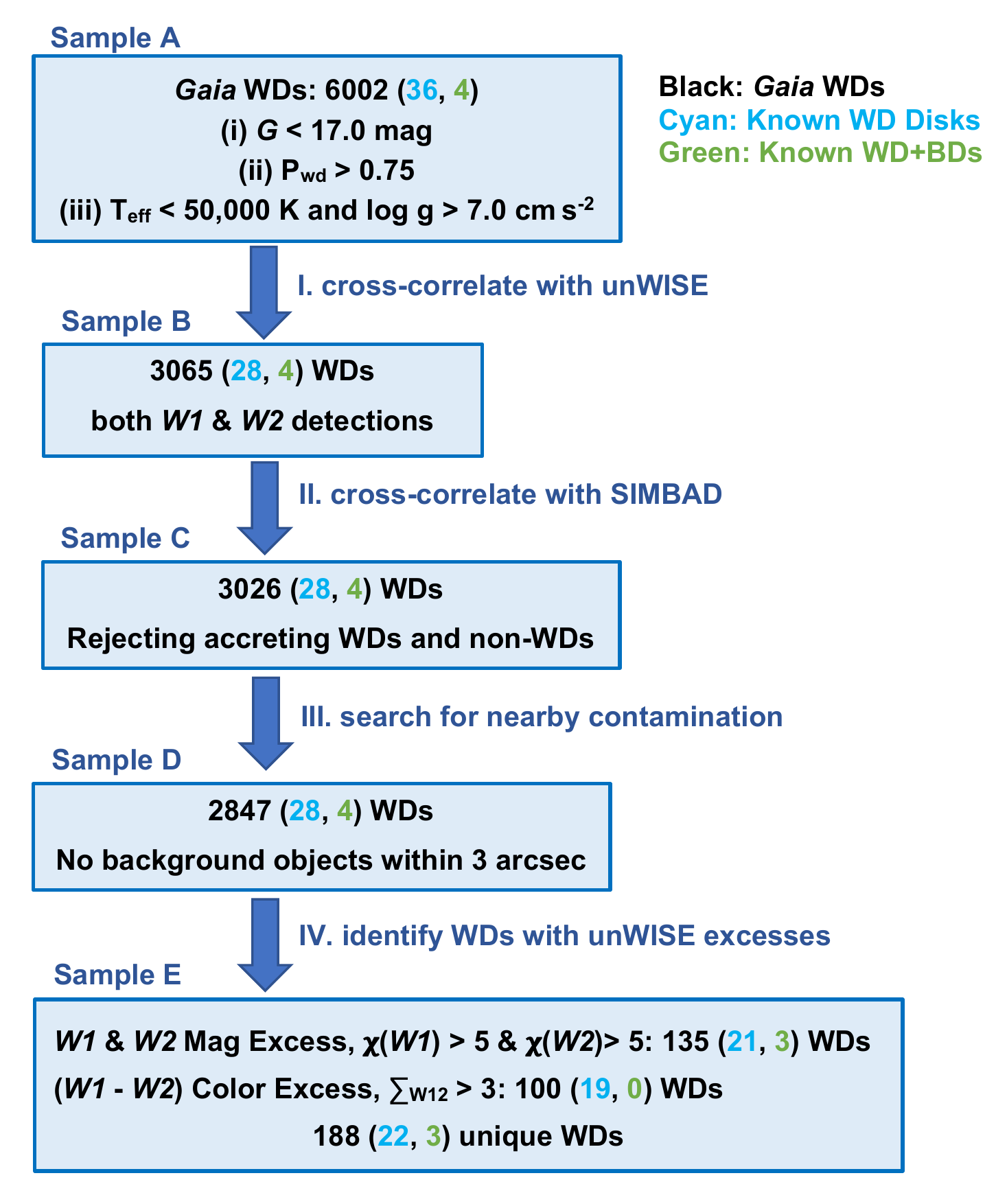}
\caption{A flowchart highlighting our selection criteria.}
\label{fig:flowchart}
\end{figure}

There are four main steps to identify infrared excess candidates, as shown in the flowchart in Figure~\ref{fig:flowchart}. We start with the sample of 6002 white dwarfs presented in section~\ref{sec:sample} (sample A) and describe each step in the following.

{\it I. Cross-correlate with unWISE}. We first cross-correlated the {\it Gaia} white dwarf sample with the unWISE catalog. Using the 2015.5 epoch {\it Gaia} coordinates and their proper motions, we calculated the coordinate of each white dwarf back to the 2014.4 epoch, which is the middle point of the unWISE catalog \citep{Schlafly2019}. The unWISE position uncertainty varies between 0{\farcs}01 for the brightest white dwarfs to 1{\farcs}5 for the faintest ones. We consider a source to have a positive detection in the unWISE catalog if the measured unWISE coordinate and the calculated coordinate using the {\it Gaia} position and proper motion agree to within 3$\sigma$. There are 3065 white dwarfs with both unWISE \textit{W1} and \textit{W2}  detections (sample B).

 {\it II. Cross-correlate with SIMBAD}. We cross-correlated Sample B with SIMBAD and found 1688 systems with an entry in the spectral type. The majority (1649) are indeed white dwarfs but 39 systems are rejected because they are either not white dwarfs (e.g., subdwarfs) or are accreting white dwarfs (e.g., cataclysmic variables). We caution that there will be other outliers and spectroscopy is needed to confirm the white dwarf nature of these objects in our sample.

{\it III. Search for nearby contamination}. {\it WISE} has a large 6{\farcs}0 beam and therefore is subject to background confusion, particularly for faint sources like white dwarfs. Therefore, we cross-correlated our sample with as many surveys as possible to search for background contamination. We started with the coordinates and proper motions from {\it Gaia DR2} (J2015.5) and calculated the expected positions in each catalog, which includes the Sloan Digital Sky Survey (SDSS) DR 12 (J2007.5), UKIRT Infrared Deep Sky Survey (UKIDSS) DR9 LAS/GCS/DXS (J2008.5), the Vista Hemisphere Survey (VHS) DR6 (J2012), the Vista Variables in the Via Lactea (VVV) DR4 (J2012), the VISTA Kilo-degree Infrared Galaxy Survey (VIKING) DR5 (J2013), the UHS (UKIRT Hemisphere Survey) DR1 (J2015), and the {\it Gaia DR2} (J2015.5). If more than one source is detected within 3{\farcs}0 of the position of the white dwarf, the unWISE photometry is considered to be contaminated and the system is excluded from the following analysis. In some cases, multiple detections of the white dwarf at different epochs could lead to multiple entries in a given catalog, mimicking background contamination. Fortunately, Gaia, SDSS and UKIDSS have generated a duplication flag when the same source is detected twice. For UHS, VIKING, VHS, and VVV, we have manually checked all the images where more than one source was detected within 1{\farcs}6 of the expected position of the white dwarf and independently confirmed the number of sources. Out of the 3065 white dwarfs in sample B, 2691 objects are detected in at least one other catalog in addition to {\it Gaia DR2} and unWISE. There are 2847 white dwarfs with no background objects within 3{\farcs}0 (Sample D, including 28 known dusty white dwarfs and four white dwarf-brown dwarf pairs), as listed in Table~\ref{tab:sample}.

\begin{deluxetable*}{lcccccccccccccccc}
\tabletypesize{\tiny}
\tablecaption{\label{tab:sample} Bright {\it Gaia} white dwarfs with reliable unWISE photometry (Sample D). RA, DEC, Exs, T$_\mathrm{eff}$ and log g are the same as those defined in Table~\ref{tab:knownir}. W1, unW1, W2, and unW2 are unWISE magnitudes and uncertainties from \citet{Schlafly2019}. $\chi$(W1) and $\chi$(W2) are magnitude excess parameters calculated from Equation~\ref{equ:flux}. $\Sigma_\mathrm{W12}$ is color excess parameter calculated from Equation~\ref{equ:color}. White dwarf mass M is from \citet{Gentile-Fusillo2019}, which is derived from evolutionary models with carbon and oxygen cores and a pure hydrogen atmosphere. SpT is the spectral type from SIMBAD and Catalog represents the catalogs that each white dwarf is detected in.
}
\tablehead{
 \colhead{RA} & \colhead{DEC} & \colhead{W1} & \colhead{unW1} & \colhead{W2} & \colhead{uncW2} & \colhead{$\mathrm{\chi}$(W1)} & \colhead{$\mathrm{\chi}$(W2)} & \colhead{$\mathrm{\Sigma_\mathrm{W12}}$}& \colhead{Exs} & \colhead{T$_\mathrm{eff}$} & \colhead{log g} & \colhead{M} & \colhead{SpT}  & \colhead{Catalog}  \\
 & & \colhead{(Mag)} & \colhead{(Mag)} & \colhead{(Mag)} & \colhead{(Mag)} && & & & \colhead{(K)} & \colhead{(cm s$^{-2}$)}& \colhead{(M$_\mathrm{\odot}$)}  
}
\startdata
 0.030471 &  29.949947 &        16.18 &      0.06 &        16.21 &      0.12 &     -1.43 &     -0.79 &         0.03 &       ... &  43346 &   7.71 &  0.55 &       DA.9 &     Gaia/SDSS/UHS/unWISE\\
 0.581904 & -52.490677 &        16.29 &      0.06 &        16.09 &      0.11 &      6.45 &      6.06 &         1.56 &     Color &  10117 &   8.07 &  0.64 &        ... &          Gaia/VHS/unWISE \\
 0.645709 &  17.455267 &        16.65 &      0.07 &        16.40 &      0.14 &     -1.38 &      1.08 &         1.62 &       ... &  21152 &   7.98 &  0.62 &      DA2.4 &     Gaia/SDSS/UHS/unWISE \\
 0.796334 & -18.366858 &        16.70 &      0.07 &        16.76 &      0.21 &     -0.78 &     -0.51 &        -0.16 &       ... &  14775 &   7.93 &  0.57 &      DA3.3 &          Gaia/VHS/unWISE \\
 0.831895 &   2.439749 &        16.42 &      0.07 &        15.96 &      0.11 &      4.51 &      6.82 &         3.14 &       ... &  13356 &   7.25 &  0.33 &        ... &  Gaia/SDSS/UKIDSS/unWISE \\
 \enddata
\tablecomments{This table is published in its entirety in the machine-readable format. A portion is shown here for guidance regarding its format and content.}
\end{deluxetable*}

\begin{figure*}
\gridline{\fig{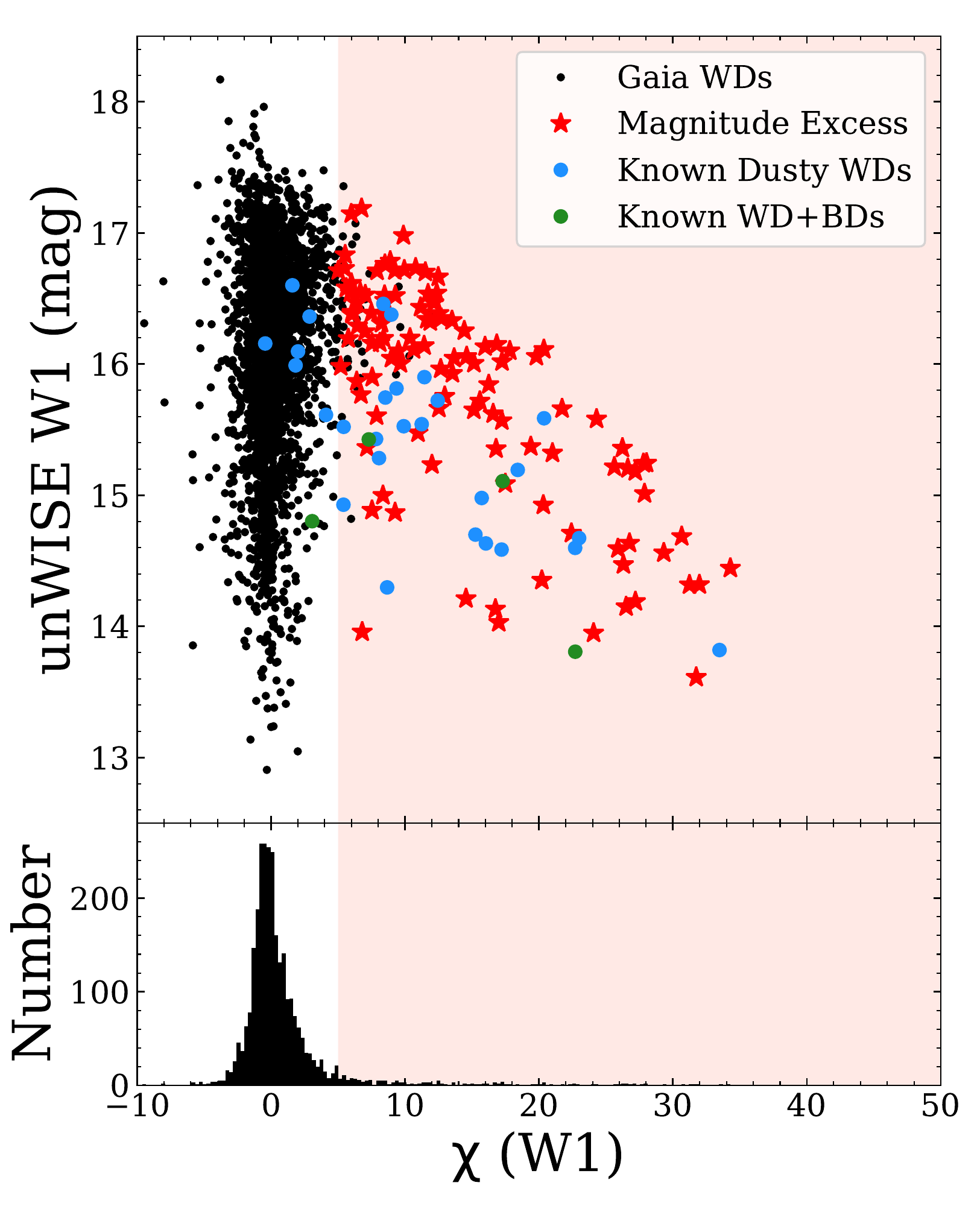}{0.333\textwidth}{}
\fig{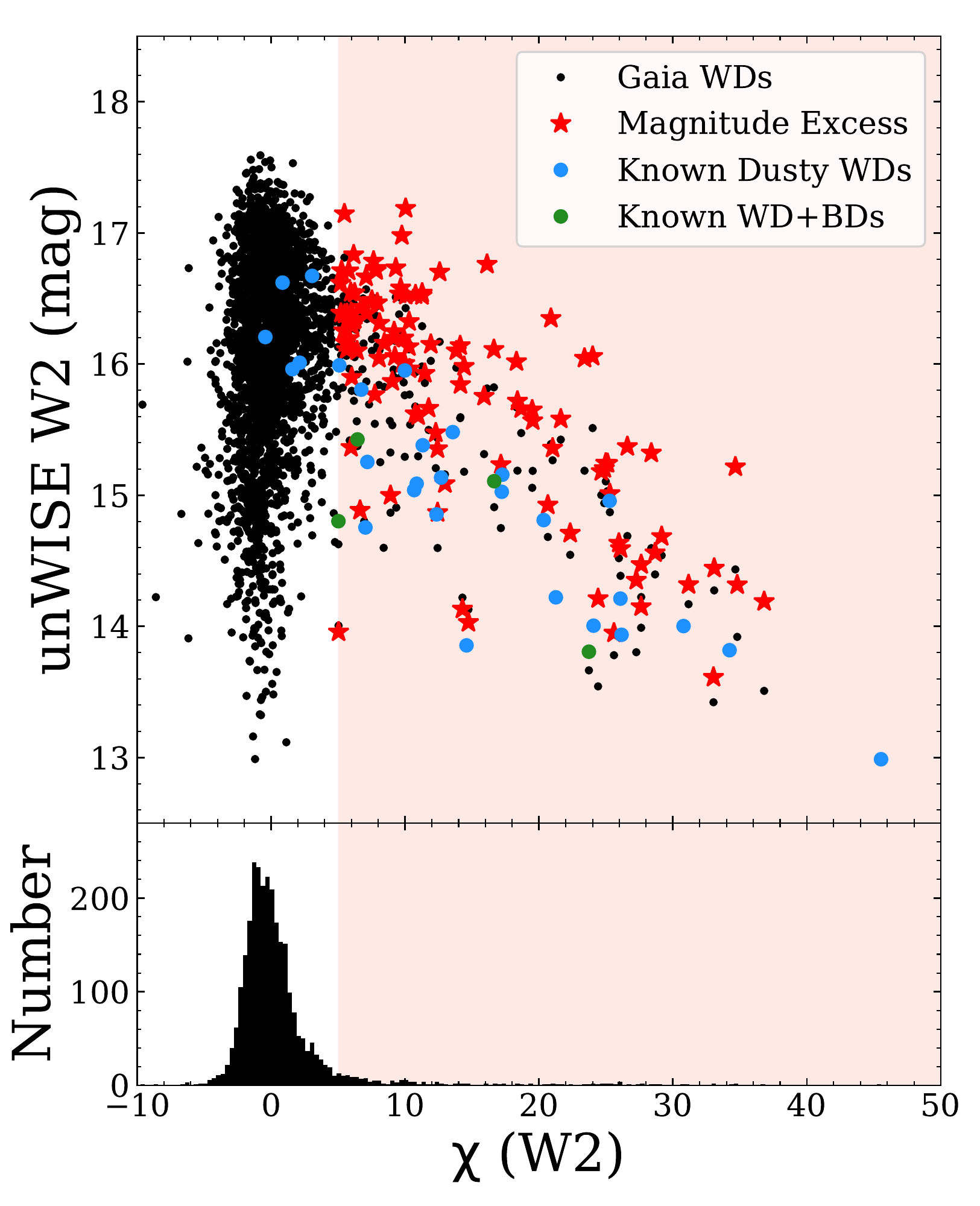}{0.333\textwidth}{}
\fig{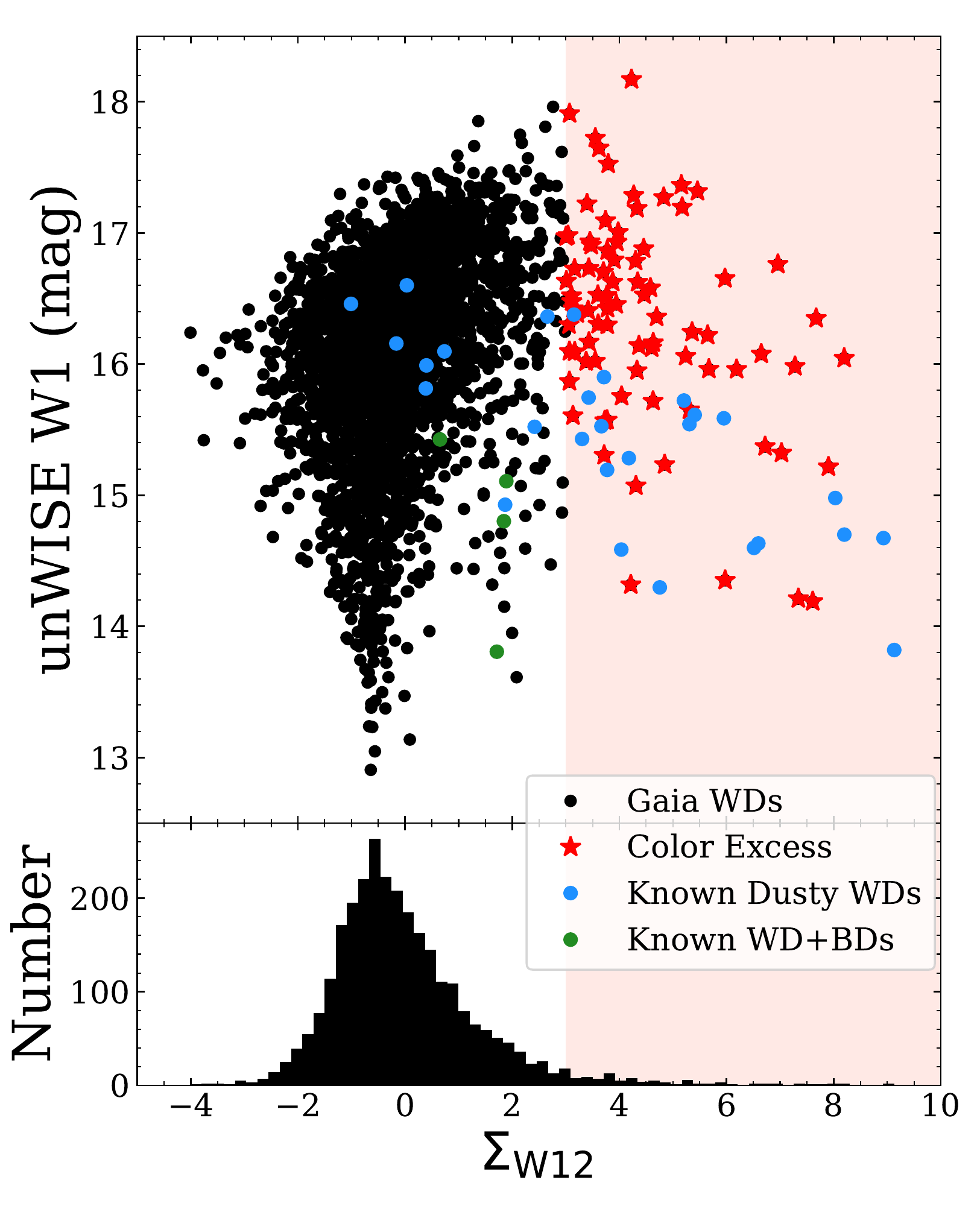}{0.333\textwidth}{}
}
\caption{White dwarfs with unWISE excesses selected from (i) the magnitude excess, which requires both $\chi$(W1) $>$ 5 and $\chi$(W2) $>$ 5 (Equation~\ref{equ:flux}, the leftmost and middle figures), (ii) the color excess, which requires $\Sigma_\mathrm{W12}$ $>$ 3 (Equation~\ref{equ:color}, the rightmost figure). The known dusty white dwarfs and white dwarf-brown dwarf pairs are also shown for comparison.
}
\label{fig:unWISEexs}
\end{figure*}

{\it IV. Identify white dwarfs with unWISE excesses}. Following the methods described in \citet{Wilson2019}, we adopted two methods to identify white dwarfs with infrared excesses, i.e., the \textit{W1}\,\&\,\textit{W2} magnitude excess and the (\textit{W1}-\textit{W2}) color excess. At a given wavelength $i$, the magnitude excess is characterized by $\chi$($i$) ,

\begin{equation}
\chi (i) = \frac{m_\mathrm{mod,i} - m_\mathrm{obs,i} }{\sqrt{\sigma_\mathrm{mod,i}^2+\sigma_\mathrm{obs,i}^2}}
\label{equ:flux}
\end{equation}
where $i$ = \textit{W1} or \textit{W2}. $m_\mathrm{obs,i}$ and $\sigma_\mathrm{obs,i}$ is the observed unWISE magnitude and uncertainty, respectively. $m_\mathrm{mod,i}$ is the calculated {\it WISE} magnitude (P. Bergeron, private communications) given the white dwarf parameters and {\it Gaia} distances. The median extinction in {\it Gaia} G band for the unWISE sample is 0.13~mag \citep{Gentile-Fusillo2019} so no correction for reddening is applied. $\sigma_\mathrm{mod,i}$ is the model uncertainty, which is taken as 5\% of the model flux. We calculated $\chi$ for both \textit{W1} and \textit{W2}, as shown in Figure~\ref{fig:unWISEexs}. Rejecting the top and bottom 10\% $\chi$ values, we find that the mean and the standard deviation is $\bar{\chi}$ (W1) = 0.17 $\pm$ 1.09, $\bar{\chi}$ (W2) = -0.05 $\pm$ 1.27. Higher $\chi$ values mean more significant infrared excesses. In this work, we consider a white dwarf to be an infrared excess candidate from the magnitude excess method if they have both $\chi$(W1) $>$ 5 and $\chi$(W2) $>$ 5.

The (\textit{W1}-\textit{W2}) color excess is characterized by $\Sigma_\mathrm{W12}$, 

\begin{equation}
\Sigma_\mathrm{W12} = \frac{m_\mathrm{obs,W1} - m_\mathrm{obs,W2} - (m_\mathrm{mod,W1} - m_\mathrm{mod,W2})}{\sqrt{\sigma_\mathrm{obs,W1}^2+\sigma_\mathrm{obs,W2}^2+\sigma_\mathrm{mod,W1}^2+\sigma_\mathrm{mod,W2}^2}}
\label{equ:color}
\end{equation}
Rejecting the top and bottom 10\% of $\Sigma_\mathrm{W12}$ values, the mean and the standard deviation is $\bar{\Sigma}_\mathrm{W12}$ = -0.13 $\pm$ 0.76. A larger $\Sigma_\mathrm{W12}$ value indicates a stronger infrared excess. Here, a system is considered to display an infrared excess from the color excess if $\Sigma_\mathrm{W12}$ $>$ 3 (see Figure~\ref{fig:unWISEexs}). There are 106 such systems, including 52 also having \textit{W1}\,\&\,\textit{W2} magnitude excess.

To be as complete as possible, we also included 179 white dwarfs that have unWISE photometry but are likely to suffer from background contamination in Table~\ref{tab:sampleB-C}. Essentially, these are the white dwarfs that are in sample C but not in sample D of Figure~\ref{fig:flowchart}. We performed the exact same kind of analysis and listed $\chi$(W1), $\chi$(W2), and $\Sigma_\mathrm{W12}$ in the table. Some of them might have a real infrared excess but the nearby source complicates the interpretation. Further investigation is needed and we exclude this sample for our following analysis.

\begin{deluxetable*}{lcccccccccccccccc}
\small
\tablecaption{\label{tab:sampleB-C} This table lists 179 white dwarfs that may have contaminated unWISE photometry. They are in sample C but not in sample D of Figure~\ref{fig:flowchart}. The columns have the same format as those in Table~\ref{tab:sample} except for the Catalog column, which lists the catalogs where background objects are detected.}
\tablehead{
 \colhead{RA} & \colhead{DEC} & \colhead{W1} & \colhead{unW1} & \colhead{W2} & \colhead{uncW2} & \colhead{$\mathrm{\chi}$(W1)} & \colhead{$\mathrm{\chi}$(W2)} & \colhead{$\mathrm{\Sigma_\mathrm{W12}}$}& \colhead{Exs} & \colhead{T$_\mathrm{eff}$} & \colhead{log g} & \colhead{M} & \colhead{SpT} & \colhead{Catalog} \\
 & & \colhead{(Mag)}  & \colhead{(Mag)} & \colhead{(Mag)} & \colhead{(Mag)} && & & & \colhead{(K)} & \colhead{(cm s$^{-2}$)}& \colhead{(M$_\mathrm{\odot}$)}  
}
\startdata
  3.134785 &  -5.559512 &     16.17 &        0.06 &     16.35 &        0.15 &      5.55 &      1.79 &        -0.90 &  ... &     10560 &             8.22 &            0.74 &  DA4.8 &     Gaia \\
  4.867334 & -21.818318 &     15.69 &        0.05 &     15.96 &        0.12 &      0.27 &     -1.73 &        -1.63 &  ... &     13415 &             7.92 &            0.56 &  DA3.7 &      VHS \\
  5.807207 &  47.883265 &     15.93 &        0.05 &     16.03 &        0.10 &      1.11 &      0.15 &        -0.48 &  ... &     19857 &             8.16 &            0.72 &    ... &      UHS \\
  7.791164 &  35.381924 &     16.59 &        0.07 &     16.42 &        0.14 &      3.75 &      3.33 &         1.03 &  ... &     11461 &             7.05 &            0.27 &    ... &     Gaia \\
 11.259384 &  -4.119719 &     16.11 &        0.06 &     16.12 &        0.12 &     -9.80 &     -6.06 &        -0.29 &  ... &      5708 &             7.52 &            0.37 &    ... &   unWISE \\
\enddata
\tablecomments{This table is published in its entirety in the machine-readable format. A portion is shown here for guidance regarding its format and content.}
\end{deluxetable*}

\section{Discussion \label{sec:discuss}}

\subsection{Caveats}

We have identified a total of 188 white dwarfs that show an infrared excess via either the magnitude excess or the color excess (sample E). Before interpreting the results, we will discuss three caveats for this analysis. 

\begin{figure}
\epsscale{1.1}
\plotone{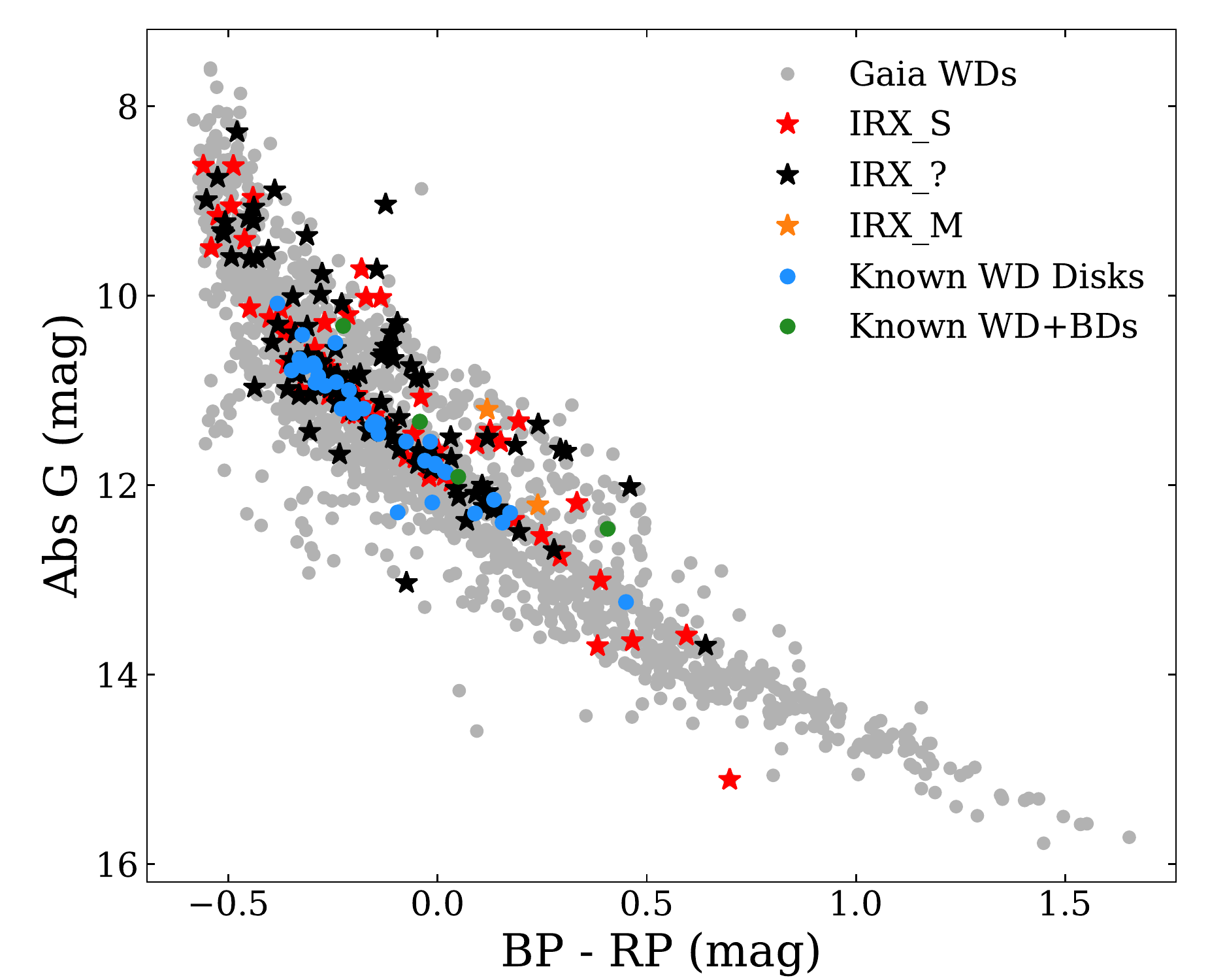}
\caption{{\it Gaia} HR diagram of white dwarfs that are spectroscopically confirmed in sample A (grey dots), infrared excess candidates in sample E (different-colored stars), known white dwarf disks (blue dots), and known white dwarf-brown dwarf pairs (green dots). Red (IRX\_S), black (IRX\_?), and orange (IRX\_M) stars represent infrared excess candidates that are spectroscopically-confirmed, not spectroscopically confirmed, and white dwarf-M dwarf pairs, respectively.}
\label{fig:GaiaHR}
\end{figure}

 {\it The starting sample are white dwarf candidates.} Our starting sample are white dwarf candidates identified using {\it Gaia} photometry and parallax \citep{Gentile-Fusillo2019}. Out of the 188 white dwarfs with an infrared excess, 83 are spectroscopically confirmed white dwarfs, including two white dwarf-M dwarf pairs. The rest 105 systems have no spectroscopic classification. Most of them are likely real white dwarfs but there will be contaminants and optical spectroscopy is needed to confirm the nature of these objects. Their positions in the {\it Gaia} Hertzsprung-Russell (HR) diagram is shown in Figure~\ref{fig:GaiaHR}.

 {\it Pure hydrogen atmosphere models are assumed for all white dwarfs.} In this analysis, the white dwarf parameters are all taken from pure hydrogen atmospheres fits reported in \citet{Gentile-Fusillo2019} and the infrared excess is identified from assuming pure hydrogen atmosphere models (P. Bergeron, private communication). Atmospheric composition can affect the white dwarf parameters \citep{Bergeron2019} and in return affect the assessment of the infrared excess. 

 {\it Background contamination is unavoidable for the infrared excess candidates.} Previous studies show that the cause of false positive for {\it WISE}-selected infrared excess objects is background contamination \citep{Barber2014,Dennihy2020}. This sample should be treated as infrared excess candidates until further confirmation.

\subsection{Magnitude Excess vs Color Excess \label{sec:fluxvcolor}}

Magnitude excess and color excess have been widely used to identify white dwarfs with an infrared excess. Essentially, magnitude excess requires the measured flux to be above the white dwarf's photosphere. This method is used in most {\it Spitzer} searches and some {\it WISE} searches of white dwarf disks \cite[e.g.,][]{Jura2007b,Farihi2009,Debes2011b}. Magnitude excess has no requirement for the shape of the infrared excess. Its accuracy is strongly dependent on white dwarf models and the flux calibrations of the observations. On the other hand, color excess requires the excess to be brighter in \textit{W2}  than \textit{W1} and it has been used in some disk searches \citep{Hoard2013, Wilson2019}. Color excess is less sensitive to the accuracy of the white dwarf models but it tends to miss objects with an unusual shape of the infrared excess.

Unresolved white dwarf-M dwarf pairs can also have {\it WISE} excesses and thousands such systems have been discovered in the SDSS \citep{Rebassa-Mansergas2016}. The frequency of M dwarf companions to white dwarfs is estimated to be 30\% \citep{Debes2011b}, much more common than white dwarf debris disks and white dwarf-brown dwarf pairs. In this study, most white dwarfs with main sequence companions are excluded because of the color cut used by \citet{Gentile-Fusillo2019} to select {\it Gaia} white dwarf candidates and the log g $>$ 7.0 cm s$^{-2}$ requirement in our initial sample. In addition, out of the 83 spectroscopically confirmed infrared excess candidates, only two are white dwarf-M dwarf pairs. Likely, the frequency of M dwarf companions is low for this sample.

Our assessments for known white dwarfs with an infrared excess are listed in Table~\ref{tab:knownir}. For the systems with unWISE detections, most dusty white dwarfs (18 out of 28) show both magnitude excess and color excess. There are three systems that only show the magnitude excess and one system that only shows the color excess. Interestingly, the three known white dwarf-brown dwarf pairs only show magnitude excess and not color excess, likely due to the methane and water absorptions in the {\it WISE} bands (see Figure~\ref{fig:sed_comp}). 

For the infrared excess candidates in Table~\ref{tab:sample}, 29 new white dwarfs (excluding the known systems) show both color excess and magnitude excess -- this is the most promising sample of disk candidates. For the following analysis, to be as complete as possible, we decided to include objects selected from either the magnitude excess or the color excess as infrared excess candidates. 

\subsection{Completeness}
In the sample of 2847 white dwarf candidates with unWISE photometry, we have identified a total of 188 white dwarfs that show {\it WISE} excesses. Apart from the 25 known objects, 14 candidates reported previously in \citet{Rebassa-Mansergas2019}, and 2 white dwarf-M dwarf pairs, the rest are all new systems that are reported here for the first time. Out of the 32 known infrared excess objects that also have unWISE photometry, our selection criteria have recovered 22 out of the 28 white dwarf debris disks and 3 out of 4 white dwarf-brown dwarf pairs. Nominally, the completeness rate of our study is 25/32 = 78\%. The seven systems that are missed by our selection criteria tend to have a weak infrared excess. For example, WD~2132+096 shows no excess at shorter wavelengths and only a 4$\sigma$ excess at 4.5~$\mu$m in {\it Spitzer} observations \citep{Bergfors2014}, while WD~2328+107 requires a very high disk inclination to fit the subtle excess \citep{Rocchetto2015}. The poorer sensitivity of {\it WISE} limits us from detecting subtle infrared excess objects. However,{\it WISE} is well suited for finding the brightest and most prominent infrared excesses around white dwarfs -- the focus of this work.

The frequency of infrared excesses around white dwarfs in this sample is 188/2847 = 6.6 $\pm$ 0.5\% assuming a Poissonian probability distribution for calculating the uncertainty. Using the 78\% completeness rate calculated above, the corrected frequency is 8.4 $\pm$ 0.6\%. We caution that this number should be treated as an upper limit because our sample is limited by source confusion \citep{Dennihy2020}. Studies using {\it Spitzer} show that the white dwarf disk fraction is about 2-4\% \citep{Barber2014,Rocchetto2015, Wilson2019}. The frequency of white dwarf-brown dwarf systems is estimated to be 0.5-2.0\% \citep{Girven2011, Steele2011}. Assuming a true infrared excess (disk and brown dwarf combined) frequency of 3\%, it implies a false positive rate of (100\% - 3\%/8.4\%) = 64\% for this sample. Even so, there would be 188 $\times$ (1 - 64\%) - 25 = 42 new white dwarfs with a real infrared excess, which will more than double the known sample. 

\begin{figure}
\epsscale{1.2}
\plotone{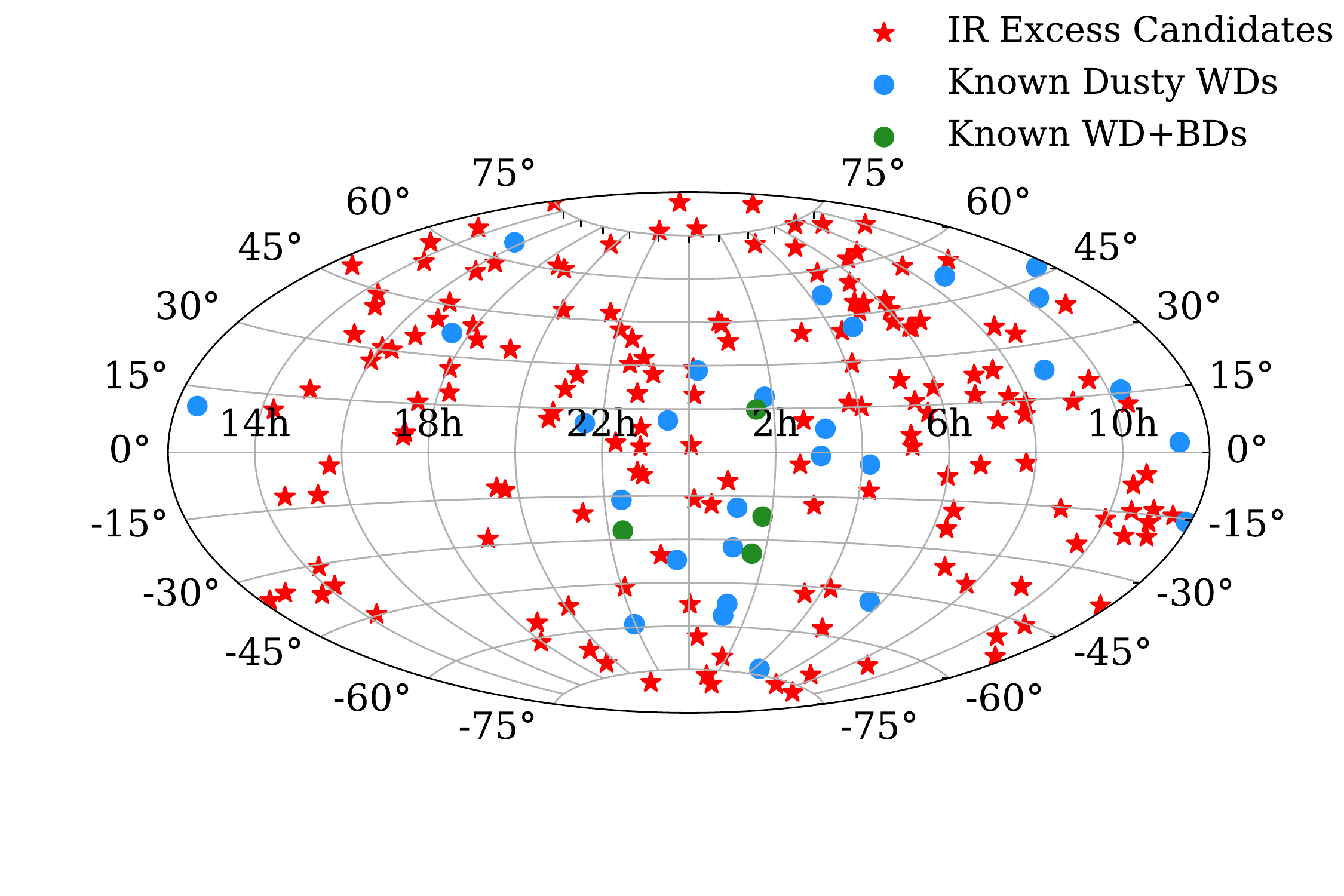}
\caption{Positions of known dusty white dwarfs, white dwarf-brown dwarf pairs, and infrared excess candidates identified in this study in the equatorial coordinate.
}
\label{fig:map}
\end{figure}

The sky positions of the infrared excess objects are shown in Figure~\ref{fig:map}. This study has significantly increased the number of infrared excess candidates. These new candidates are evenly scattered all over the sky except for areas around the Milky Way disk, where the stellar density increases significantly and the unWISE sensitivity becomes a lot worse.

\subsection{Properties of The Infrared Excess Candidates \label{sec:properties}}

\begin{figure*}
\gridline{\fig{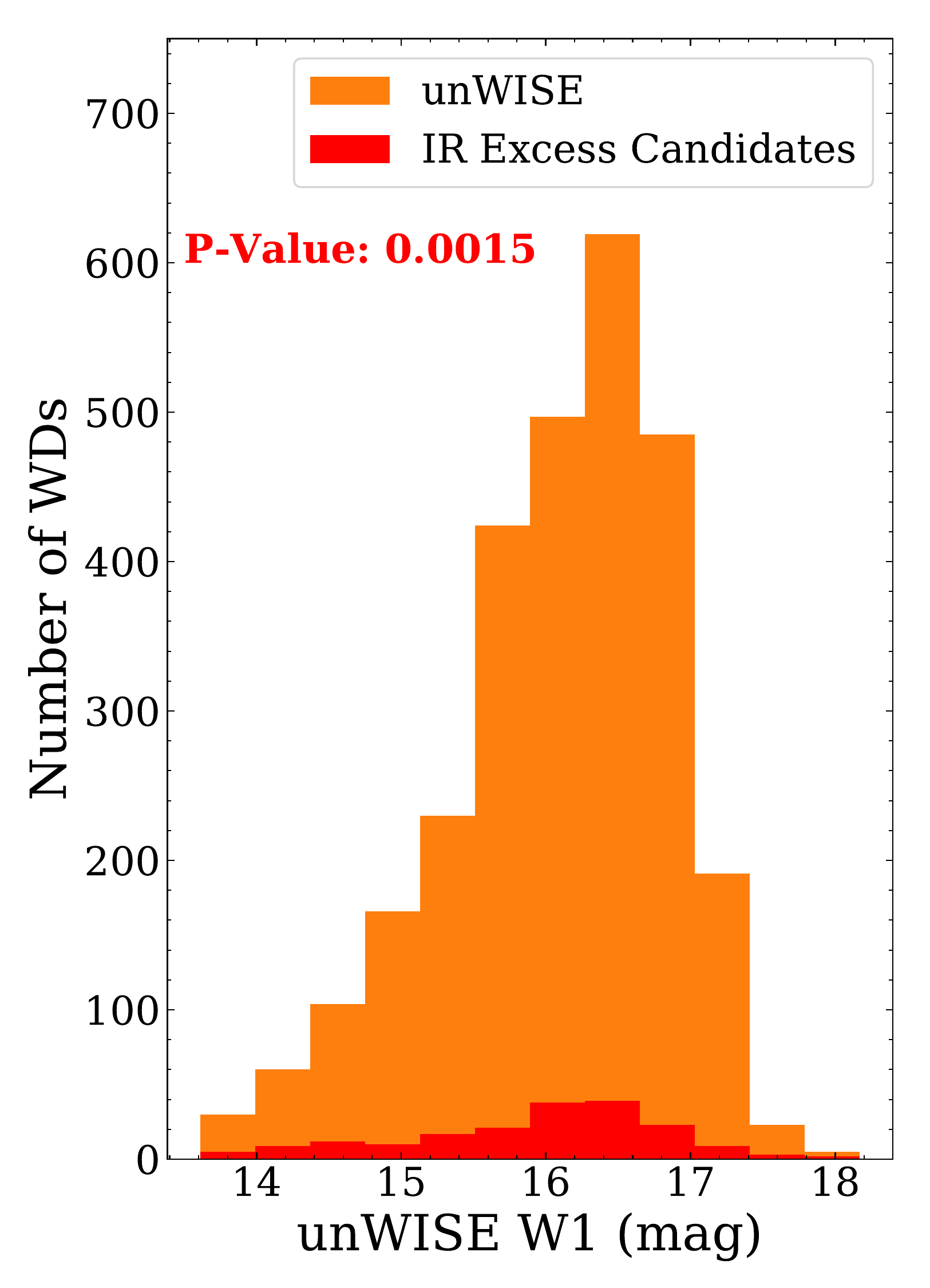}{0.25\textwidth}{}
\fig{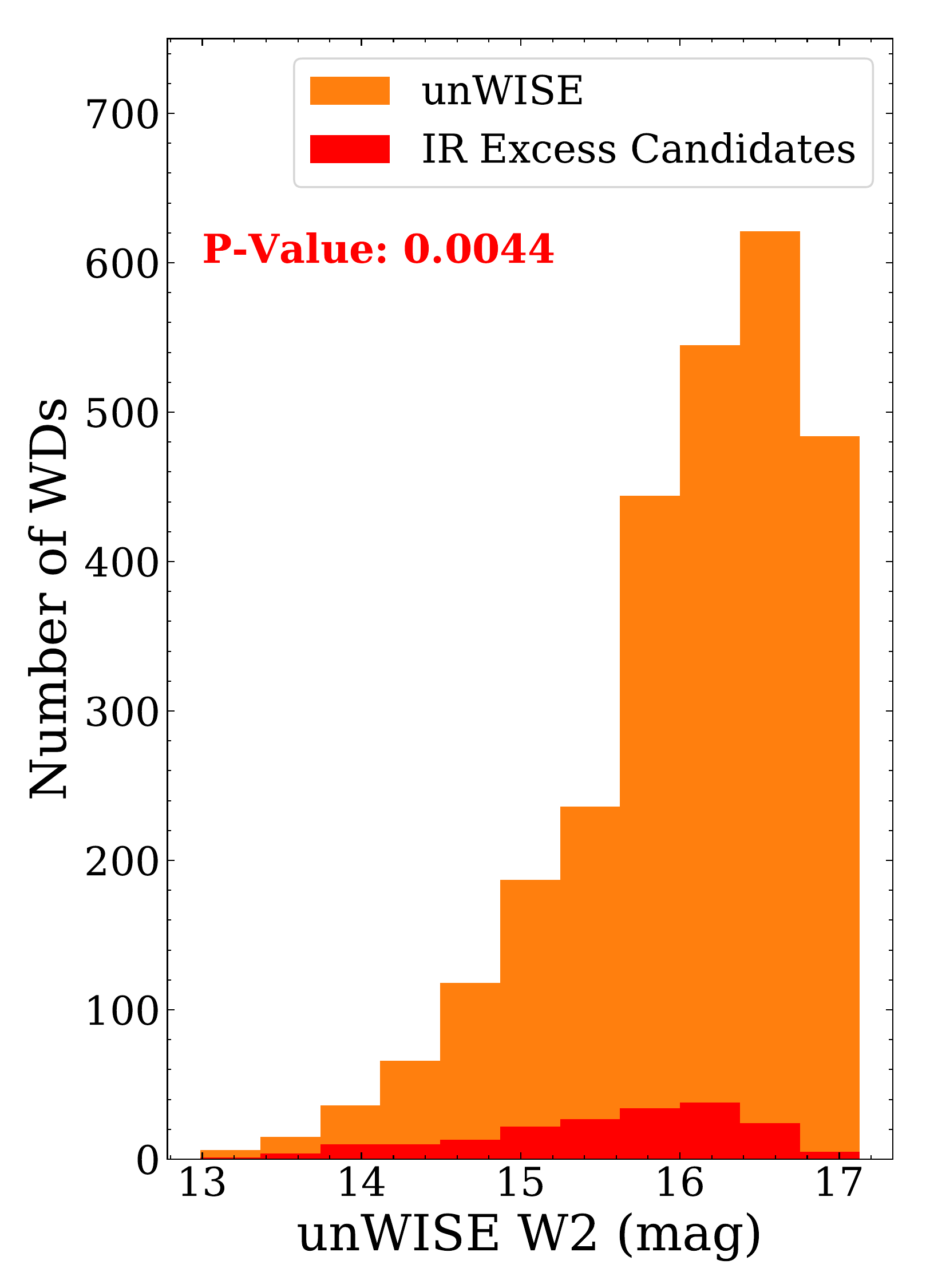}{0.25\textwidth}{}
\fig{Fig7_T.pdf}{0.25\textwidth}{}
\fig{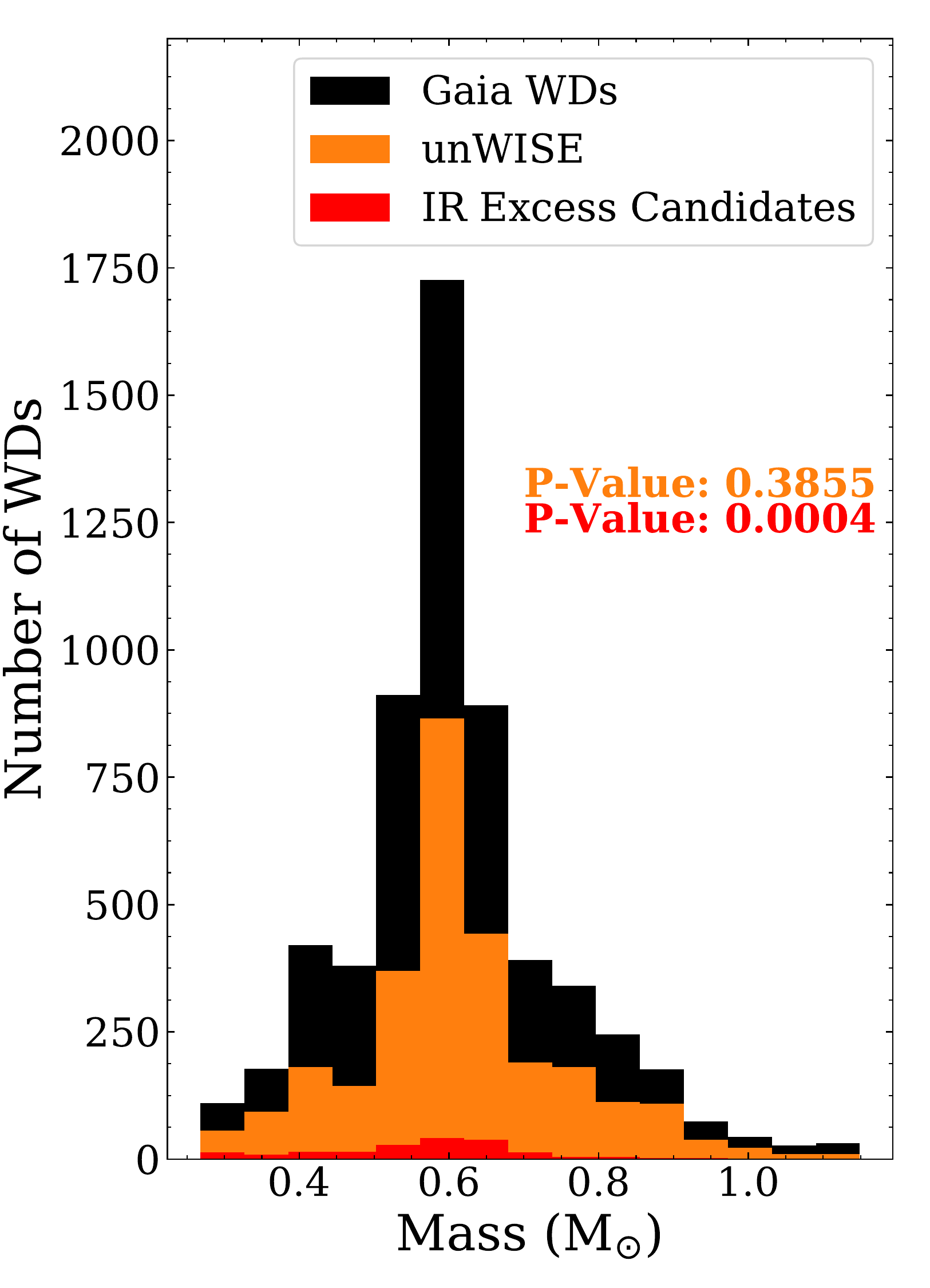}{0.25\textwidth}{}
}
\caption{Number of white dwarfs for the {\it Gaia} sample (sample A in Figure~\ref{fig:flowchart}, black), the unWISE sample (sample D in Figure~\ref{fig:flowchart}, orange), and the infrared excess candidate sample (sample E in Figure~\ref{fig:flowchart}, red) as a function of the unWISE W1 and W2 magnitude, the white dwarf temperature, and the mass, respectively. In each panel, we list the p-values from the Kolmogorov-Smirnov test between the {\it Gaia} sample and the unWISE sample in orange and the unWISE sample and the infrared excess candidate sample in red. In each panel, the distributions are not significantly different.
}
\label{fig:frequency}
\end{figure*}

Now we assess the overall properties of the {\it Gaia} white dwarf sample (sample A), the unWISE sample (sample D), and the infrared excess candidate sample (sample E). A comparison of these three distributions as a function of the unWISE magnitude, the white dwarf temperature, and the white dwarf mass is shown in Figure~\ref{fig:frequency}. We performed Kolmogorov-Smirnov tests between these distributions and did not find any significant differences, particularly between the unWISE sample and the infrared excess candidate sample. 

{\it The unWISE magnitude distribution}: The P-values for both unWISE W1 and W2 are very small between the unWISE sample and the infrared excess candidate sample. It shows that generally our selection method is not biased against the brightness of a white dwarf in unWISE. However, we can see in Figure~\ref{fig:unWISEexs} that the scatters in $\chi$ and $\Sigma$ become larger for fainter white dwarfs. Extra caution is needed to interpret the origin of the infrared excess at the faint end.

{\it The temperature distribution}: At the cool end (T$_\mathrm{eff}$ $<$ 10,000~K), we have identified 21 infrared excess candidates. This is contrary to previous findings that there is a lack of dust disks around cool white dwarfs \citep{XuJura2012}. At the hot end, there are 23 white dwarfs with a temperature higher than 25,000~K, which is the hottest known white dwarf with an infrared excess (PG~0010+280, Table~\ref{tab:knownir}). At these high temperatures, dust particles are expected to fully sublimate within the tidal radius of the white dwarf \citep{vonHippel2007}. The origins of the infrared excesses around those cool (T$_\mathrm{eff}$ $<$ 10,000~K) and hot white dwarfs (T$_\mathrm{eff}$ $>$ 25,000~K) deserve further investigation.

{\it The mass distribution}: There is little difference in the mass distributions between the unWISE sample and the infrared excess candidate sample. We have identified 11 white dwarfs with with a mass $\lesssim$ 0.3 M$_\mathrm{\odot}$, which are called extremely low mass (ELM) white dwarfs and are often in binaries \citep{Brown2020}. It is very unusual to see a strong infrared excess around an ELM white dwarf and follow-up study is needed to understand its origin. In addition, there are 14 disk candidates around white dwarfs with a mass larger than 0.75~M$_\mathrm{\odot}$. However, most of these white dwarfs are not spectroscopically confirmed and the white dwarf parameters could be different depending on the atmospheric composition \citep{Bergeron2019}. If the infrared excess is confirmed, it can help us understand the frequency of planetary systems around massive stars which is otherwise hard to constrain \citep{Barber2016}.

\section{Conclusions \label{sec:conclusion}}

In this paper, we searched for infrared excesses around 2847 bright {\it Gaia} white dwarf candidates (G $<$ 17.0 mag) using the unWISE catalog. We compared the performance of unWISE with ALLWISE and found that unWISE generally reports a smaller uncertainty and more reliable photometry for faint objects, making it a very useful catalog for studying white dwarfs. We explored two methods to identify infrared excesses -- the \textit{W1}\,\&\,\textit{W2} magnitude excess and the (\textit{W1}-\textit{W2}) color excess. Using either method, a total of 188 infrared excess candidates are identified, including 22 known white dwarf debris disks and three known white dwarf-brown dwarf pairs. The nominal completeness rate is 78\%. There are seven known systems missed by our criteria because their infrared excesses are subtle and only detected by {\it Spitzer}. In addition, we did not find any correlation between the presence of an infrared excess and a white dwarf's temperature or mass. 

We caution that without additional confirmation, this sample should be treated as infrared excess candidates because background confusion is unavoidable due to the large beam size of {\it WISE}. If the infrared excess is real, it could either be a debris disk or a low-mass companion as their SEDs are very similar between 1--5~$\mu$m. Disentangling between these two scenarios and further characterization of this sample will be presented in future works.

\smallskip
{\it Acknowledgements.}  We thank the anonymous referee for helpful suggestions that improved the quality of the paper. We thank E. Schlafly and D. Lang for helpful discussions on the applications of the unWISE catalog. We also would like to thank P. Bergeron for sharing the white dwarf model calculations. We are grateful for discussions on the content of the paper with A. Rebassa-Mansergas, A. Nitta, N. Hallakoun, B. Klein, S. Kleinman, and S. Leggett. 

This work presents results from the European Space Agency space mission {\it Gaia} and data products from the Wide-field Infrared Survey Explorer, which is a joint project of the University of California, Los Angeles, and the Jet Propulsion Laboratory/California Institute of Technology, funded by the National Aeronautics and Space Administration. This work also use data from the UKIRT Infrared Deep Sky Survey, Vista Hemisphere Survey, Two Micron All-sky Survey, the Sloan Digital Sky Survey, and Pan-STARRS.

This work is supported by the international Gemini Observatory, a program of NSF's NOIRLab, which is managed by the Association of Universities for Research in Astronomy (AURA) under a cooperative agreement with the National Science Foundation, on behalf of the Gemini partnership of Argentina, Brazil, Canada, Chile, the Republic of Korea, and the United States of America. This work is also partly supported by the Heising-Simons Foundation via the 2019 Scialog program on Time Domain Astrophysics.

\end{CJK}

\software{Astropy \citep{Astropy2013,Astropy2018}, Scipy \citep{Scipy}, Matplotlib \citep{Matplotlib}, Pandas \citep{Pandas}, TOPCAT \citep{TOPCAT}}


\end{document}